\documentclass[
twocolumn,
]{ceurart}

\usepackage{listings}

\usepackage{amsmath}
\usepackage{amssymb}
\usepackage{amsthm}
\usepackage{multirow}
\usepackage{adjustbox}
\usepackage{booktabs}
\usepackage{enumitem}
\usepackage{algorithm}
\usepackage{algorithmic}
\usepackage{tabularx}
\usepackage{graphicx}
\usepackage{subcaption}
\usepackage{float}
\usepackage[utf8]{inputenc}
\usepackage[switch]{lineno}

\begin{document}

\copyrightyear{2024}
\copyrightclause{Copyright for this paper by its authors.
  Use permitted under Creative Commons License Attribution 4.0
  International (CC BY 4.0).}

\conference{, Sep 2--3, 2024, Girona, Spain}
\title{Instance Configuration for Sustainable Job Shop Scheduling}


\author[1,2]{Christian Pérez}[orcid=0000-0002-9121-7939,email=cripeber@upv.es,url=https://gps.blogs.upv.es/,]
\cormark[1]
\fnmark[1]
\author[1]{Carlos March}[orcid=0009-0009-7525-9133,email=cmarmoy@upv.es,url=https://gps.blogs.upv.es/,]
\fnmark[1]
\author[1]{Miguel Salido}[orcid=0000-0002-4835-4057,email=msalido@upv.es,url=https://gps.blogs.upv.es/,]
\address[1]{Universitat Politècnica de València, Instituto de Automática e Informática Industrial,
  Camì de Vera S/N, Valencia, Spain}
\address[2]{Universitat Politècnica de València, valgrAI - Valencian Graduate School and Research Network of Artificial Intelligence, Camì de Vera S/N, Valencia, Spain}

\cortext[1]{Corresponding author.}
\fntext[1]{These authors contributed equally.}

\begin{abstract}
The Job Shop Scheduling Problem (JSP) is a pivotal challenge in operations research and is essential for evaluating the effectiveness and performance of scheduling algorithms. Scheduling problems are a crucial domain in combinatorial optimization, where resources (machines) are allocated to job tasks to minimize the completion time (makespan) alongside other objectives like energy consumption. This research delves into the intricacies of JSP, focusing on optimizing performance metrics and minimizing energy consumption while considering various constraints such as deadlines and release dates. Recognizing the multi-dimensional nature of benchmarking in JSP, this study underscores the significance of reference libraries and datasets like JSPLIB in enriching algorithm evaluation. The research highlights the importance of problem instance characteristics, including job and machine numbers, processing times, and machine availability, emphasizing the complexities introduced by energy consumption considerations.

An innovative instance configurator is proposed, equipped with parameters such as the number of jobs, machines, tasks, and speeds, alongside distributions for processing times and energy consumption. The generated instances encompass various configurations, reflecting real-world scenarios and operational constraints. These instances facilitate comprehensive benchmarking and evaluation of scheduling algorithms, particularly in contexts of energy efficiency. A comprehensive set of 500 test instances has been generated and made publicly available, promoting further research and benchmarking in JSP. These instances enable robust analyses and foster collaboration in developing advanced, energy-efficient scheduling solutions by providing diverse scenarios.
\end{abstract}

\maketitle

\section{Introduction}

The Job Shop Scheduling Problem (JSP) stands as a cornerstone in the realm of operations research and optimization, representing a fundamental challenge pivotal for evaluating algorithmic effectiveness and performance. In essence, JSP revolves around the intricate task of allocating jobs to machines within a manufacturing environment to optimize a plethora of performance metrics, ranging from makespan and flow time to tardiness, resource utilization, and energy consumption \cite{xiong2022survey}. The process of benchmarking in JSP is multi-dimensional, necessitating the definition and evaluation of metrics such as makespan, energy consumption, and tardiness to assess scheduling efficiency and resource utilization \cite{fontes2023hybrid}. Reference libraries and datasets like JSPLIB play an indispensable role in these benchmarking endeavours, furnishing researchers with a rich array of instances sourced from seminal works and experimental studies, thereby enriching the evaluation of algorithms \cite{torres2014psplib}.

A profound understanding of problem instance characteristics significantly shapes benchmarking efforts in JSP. Factors such as the number of jobs and machines, variability in processing times, machine availability, and precedence relationships exert notable influences on algorithm performance \cite{el_kholany2022problem}. Furthermore, incorporating energy consumption considerations, contingent upon machine speed and operational attributes, introduces an added layer of complexity to these instances \cite{perez2020metaheuristic}. The delicate balance between energy consumption and scheduling decisions emerges as paramount in achieving energy efficiency goals without compromising production targets \cite{dai2019multi}.

In recent years, the spotlight on addressing energy efficiency within the realm of JSP has intensified, driven by its profound environmental and economic implications \cite{perez2023hybrid}. Strategies involving integrating speed-adjustable machines and vehicles have been explored as avenues to optimize energy consumption while upholding productivity levels \cite{salido2016genetic}. Concurrently, developing advanced algorithms and optimization techniques tailored to tackle energy-related challenges has seen significant advancements, considering factors such as machine speed, idle time, and energy requirements \cite{jyothi2023minimizing}. Real-world implementations of these energy-efficient strategies have yielded tangible benefits, manifesting in substantial cost savings and positive environmental impacts \cite{ham2021energy}.

In conclusion, the imperative of energy efficiency in JSP research has become increasingly pronounced, paralleling the traditional focus on performance metrics \cite{kotary2022fast}. Benchmarking is a linchpin in evaluating the efficacy of energy-efficient scheduling strategies, furnishing invaluable insights into their ramifications on production efficiency and energy consumption \cite{toma2024solving}. Manufacturers can embark toward more sustainable and economically viable operations by harnessing advanced optimization techniques and leveraging real-world implementations.

\section{Instance Configuration}
Test sets play a vital role in JSP research by providing a standardized platform for comparing algorithmic approaches. Their diversity enables researchers to evaluate various algorithms, from heuristics to exact methods, identifying strengths, weaknesses, and potential limitations across different contexts \cite{hoorn2017current}. However, as industrial systems evolve, the complexity of real-world problems increases, necessitating the development or expansion of test sets to simulate real-world scenarios better.

To address this need, the proposed instance configurator operates with the following parameters:

\begin{itemize}[leftmargin=*]
	\item $J=\{0,\dots,n\}$: the set of jobs.
	\item $M=\{0,\dots,m\}$: the set of machines.
	\item $S=\{0,\dots,s\}$: the set of speeds, indexed by $ s $ in $ S $.
	\item $T_{j}$: the set of tasks in job $j$, indexed by $t_{jt} \in T_j ,~ \forall j\in J, ~\forall t \in M$ .
	\item $D_{jt}$: the due date of task job $t_{jt} ~ \forall j\in J,~ \forall t\in M$.
	\item $R_{jt}$: the release date of task job $t_{jt} ~ \forall j\in J,~ \forall t\in M$.
	\item $P_{jts}$: the processing time of task job $t_{jt},~ \forall j\in J ,~ \forall t\in M,~ \forall s \in S $.
	\item $E_{jts}$: the energy consumption for processing task job $t_{jt},~ \forall j\in J ,~ \forall t\in M,~ \forall s \in S $.
\end{itemize}

\begin{algorithm}[hbtp]
    \caption{Instance Configurator}
    \textbf{input: } Quantity of instances $Q$, Number of machines $M$, Number of jobs $J$,  Number of tasks $T$,  Release and due date type $rrdd$, Random seed $seed$, Distribution $dist$\\
    \textbf{output: } Generated instances $G$
    \begin{algorithmic}[1]
        \STATE $SetSeed(seed)$
        \STATE $G \leftarrow [~~]$ \label{alg:G}
        \FOR{$q$ from $1$ to $Q$}
            \STATE $O \leftarrow GenerateJobsOperations(T, J, M)$  \label{alg:T}
            \STATE $P \leftarrow GenerateProcessingTimes(J, M, S)$ \label{alg:P}
            \STATE $E \leftarrow GenerateEnergy(J, M, S)$ \label{alg:E}
            \STATE $R,D\leftarrow GenerateRDDate(J, M, dist)$ \label{alg:RD}
            \STATE $G \leftarrow G \cup JSP(O, P, E, R, D)$ \label{alg:JSP}
        \ENDFOR
        \RETURN $G$
    \end{algorithmic}
    \label{alg:generator}
\end{algorithm}

The process of configuring instances, managed by Algorithm \ref{alg:generator}, is initiated by receiving several input parameters: the number of instances to generate $Q$, the number of machines $M$, the count of jobs $J$, the number of tasks $t$, the types of release and due dates $rrdd$, random seeds $seeds$, and the distribution $dist$.

Subsequently, the algorithm executes a series of steps to generate instances systematically. The random seed is initialized to ensure reproducibility, and an empty list $G$ is created to store the generated instances (Line \ref{alg:G}). The algorithm generates jobs and tasks within a loop iterating through each instance $q$ from $0$ to $Q - 1$ (Line \ref{alg:T}).

\begin{figure*}[hbtp]
    \centering
    \includegraphics[width=0.7\linewidth]{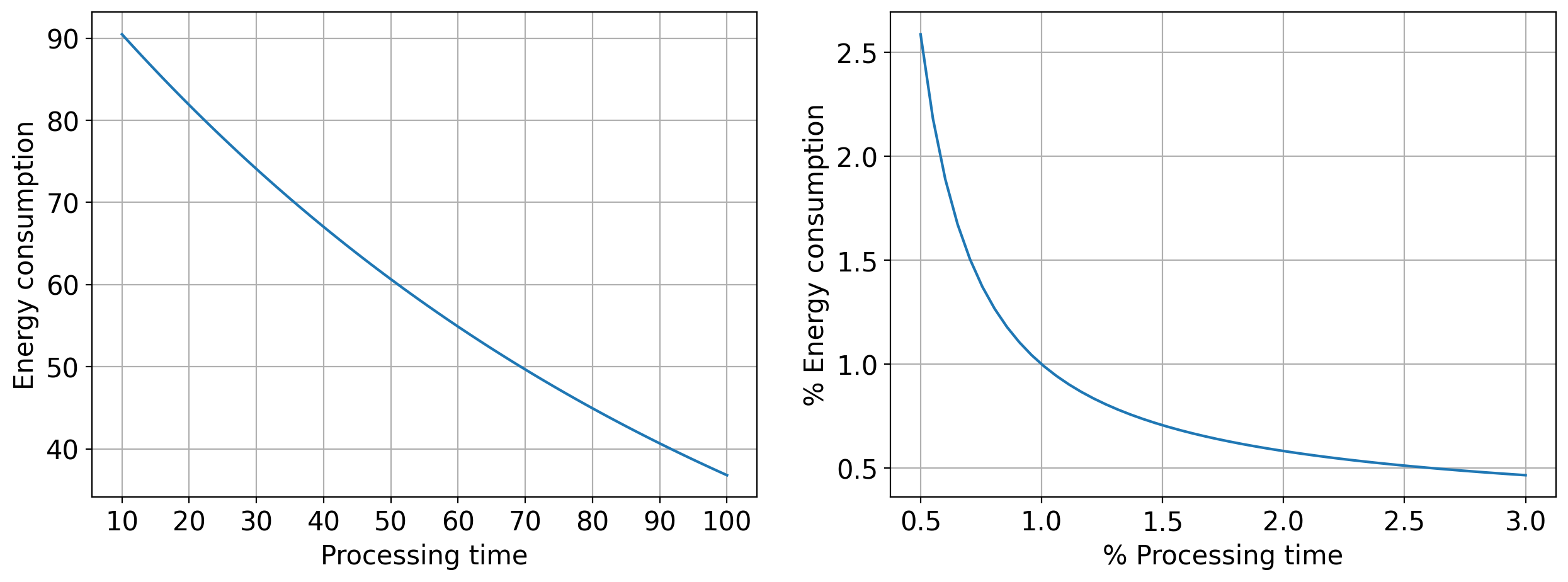}
    \caption{Distribution of processing times and the relationship between time and energy}
    \label{fig:FandG}
\end{figure*}

Next, each job operation's processing times and energy consumption are generated.
\begin{align}
    \label{eq:fx}
    f(x) = \lfloor e^{-\frac{x}{100}} \times 100 \rfloor 
\end{align}
\begin{align}
    \label{eq:gx}
    g(x) = 4.0704 \times \frac{\log(2)}{\log(1 + (x \times 2.5093)^3)}
\end{align}

The functions responsible for generating the processing times (line \ref{alg:P}) and energy consumption (line \ref{alg:E}) for the combination of jobs, machines, and speeds are managed by functions $f(x)$ \ref{eq:fx} and $g(x)$ \ref{eq:gx}, as illustrated in Figure \ref{fig:FandG} as studied in \cite{morillo_torres2014psplib}. These functions balance the variables to establish a correlation between processing time and energy consumption. In these equations, $x$ represents the processing time for Equation $f(x)$ \ref{eq:fx} and the energy consumption for Equation $g(x)$ \ref{eq:gx}. Speeds are generated by obtaining the energy consumption percentage for each speed using Equation $f(x)$ \ref{eq:fx}, which models an inverse relationship between processing time and energy consumption. This involves dividing the interval [0.5, 3] into $|S|-1$ equal parts, where the boundaries of these new intervals correspond to the energy consumption percentages for each speed. Subsequently, Equation $g(x)$ \ref{eq:gx} is utilized to determine the fraction of time corresponding to each speed.

Release and due dates are computed using functions based on the chosen distribution (Line \ref{alg:RD}). Each distribution offers distinct characteristics suited for modeling various real-world scenarios. 

The exponential distribution, defined by its probability density function $ f(x;\lambda) = \lambda e^{-\lambda x} $, is ideal for modeling the time until an event occurs, such as machine failures or job arrivals, assuming a constant hazard rate $ \lambda > 0 $. Its mean is $ \frac{1}{\lambda} $. The Gaussian distribution, with mean $ \mu $ and standard deviation $ \sigma $, has a probability density function $ f(x;\mu,\sigma) = \frac{1}{\sigma \sqrt{2\pi}} e^{-\frac{(x-\mu)^2}{2\sigma^2}} $. This distribution represents naturally occurring variations in processing times or delays. The uniform distribution generates values evenly within a specified range, defined by the probability density function $ f(x;a,b) = \frac{1}{b - a} $, where $ a $ and $ b $ are the lower and upper bounds, respectively. It provides a straightforward way to explore a range of scenarios without bias towards any particular value. These distributions were selected due to their unique properties and common use in modeling different real-world data types, enhancing the diversity and comprehensiveness of the generated instances \cite{gui2024domain}.

Additionally, a random start is chosen for each job within a specified range for the release and due date intervals. The time interval between release and due dates is determined based on the median processing time. A random value within the corresponding interval is generated depending on the chosen distribution. This comprehensive approach ensures the creation of instances encompassing a wide range of scenarios, facilitating robust analyses and evaluations.

These steps culminate in constructing a JSP instance (Line \ref{alg:JSP}), incorporating the generated data. Finally, the algorithm returns the list $G$ containing the generated instances. This systematic approach ensures the creation of instances that cover diverse scenarios, which is crucial for comprehensive analyses and evaluations.

\section{Generated Problems}
A comprehensive set of random instances has been generated following the procedure described in Algorithm \ref{alg:generator}. These instances exhibit diverse characteristics: the number of jobs ranges from thirty to two hundred fifty, and the number of machines ranges from three to twenty. Normal, exponential, and uniform distributions were utilized in the generation process.

Each instance was extended by relaxing release and due date restrictions. Leveraging three different speed scales, variations of each problem were created, maintaining identical data but with different operational speeds. Specifically, two additional instances were derived from each original: one incorporating the first, third, and fifth-speed scaling and another utilizing only the third-speed scaling.

In addition to varying job and machine counts, the instances encompass different operational complexities and constraints. For instance, some involve jobs with precedence constraints, necessitating certain jobs to be finished before others begin. This complexity challenges algorithms to find optimal solutions efficiently. The chosen distributions—normal, exponential, and uniform—offer a spectrum of scenarios, ranging from predictable and evenly spread job times to highly variable and unpredictable duration. This diversity ensures that the generated instances serve as robust benchmarks to evaluate the performance of scheduling algorithms under varied conditions.

Moreover, a collection of 500 test instances has been generated and made publicly accessible through \cite{planning_and_scheduling_research_group_2024_11640131}. These instances incorporate mixed distributions and speed scalings, providing researchers with a comprehensive dataset to evaluate the efficacy of scheduling algorithms. The research group aims to foster collaboration and innovation in planning and scheduling research by facilitating access to these instances. Researchers can use these standardized problems to compare methods and contribute to advancing scheduling solutions.

In summary, the generated instances cover a wide range of job and machine configurations, distribution types, and speed variations, making them suitable for diverse scheduling and planning research applications. Their availability for public use enhances their utility, promoting collaboration and enabling continuous improvement and benchmarking in the field.

\section{Acknowledgments}
The authors gratefully acknowledge the financial support of the European Social Fund (Investing In Your Future), the Spanish Ministry of Science (project  PID2021-125919NB-I00), and valgrAI - Valencian Graduate School and Research Network of Artificial Intelligence and the Generalitat Valenciana, and co-funded by the European Union.
\bibliography{bib}

\end{document}